\documentclass[prd,showpacs,preprintnumbers,amsmath,amssymb]{revtex4}

\usepackage{dcolumn}
\usepackage{bm}
\usepackage{amsmath,amssymb,epsfig,float}

\begin{document}

\title{Quantum Entanglement and Teleportation in Higher Dimensional
Black Hole Spacetimes}

\author{Xian-Hui Ge}\email{gexh@apctp.org}
\affiliation{Asia Pacific Center for Theoretical Physics, Pohang
790-784, Republic of Korea}
\author{Sang Pyo Kim}\email{sangkim@kunsan.ac.kr}
\affiliation{Department of Physics, Kunsan National University,
Kunsan 573-701, Korea\\Asia Pacific Center for Theoretical Physics,
Pohang 790-784, Korea}

\date{\today}
\begin{abstract}
We study the properties of quantum entanglement and teleportation in
the background of stationary and rotating curved space-times with
extra dimensions. We show that a maximally entangled Bell state in
an inertial frame becomes less entangled in curved space due to the
well-known Hawking-Unruh effect. The degree of entanglement is found
to be degraded with increasing the extra dimensions. For a finite
black hole surface gravity, the observer may choose higher frequency
mode to keep high level entanglement. The fidelity of quantum
teleportation is also reduced because of the Hawking-Unruh effect.
We discuss the fidelity as a function of extra dimensions, mode
frequency, black hole mass and black hole angular momentum parameter
for both bosonic and fermionic resources.
\end{abstract}

\maketitle

\section{Introduction}

The new field of quantum information has made rapid progresses in
recent years. As relativistic field theory provides not only a more
complete theoretical framework but also many experimental setups,
relativistic quantum information theory may become an essential
theory in the near future, with possible applications to quantum
teleportation. Quantum entanglement has already been studied in
relativistic frames, inertial or not
\cite{peres,cza,ging,alsing,mil,al2,p2,berg,mann,ge,ge2}. Czachor
\cite{cza} studied a version in which an electron is in paramagnetic
resonance with relativistic particles, and Peres et al demonstrated
that the spin of an electron is not covariant under Lorentz
transformation \cite{p2}. Moreover, Alsing and Milburn \cite{alsing}
studied the effect of Lorentz transformation on maximally
spin-entangled Bell states in momentum eigenstates and Gingrich and
Adami \cite{ging} derived a general transformation rule for the
spin-momentum entanglement of two qubits. The recent work of Alsing
and Milburn extended the result to a situation where one observer is
accelerated \cite{al2}. Fuentes-Schuller and Mann calculated the
entanglement between two free modes as seen by an inertial observer
detecting one mode and a uniformly accelerated observer detecting
the other mode \cite{mann}. Recently, one of us (X.-H.G.) discussed
a possible extension to the gravitational field of the quantum
teleportation in four dimensional spacetime \cite{ge,ge2}.

Quantum information theory has not generally been regarded as a
theory of spacetime, and entanglement and teleportation has usually
been discussed only in four dimensional spacetime. However, if our
spacetime is not the usual four dimensional one, the structure of
spacetime, in particular, the extra dimensions may still influence
the nature of entanglement. In the last decade, string theory with
an extra space compactified at a larger length scale or lower energy
scale than the Planck scale has been an attractive idea to solve the
gauge hierarchy problem and possibly a candidate for quantum
gravity\cite{arkani}. The interest on extra dimensions has escalated
from an expectation that in the coming years Large Hardon Collider
(LHC) at CERN may create mini-black holes and thus signal the
effects of extra dimensions because of the relevant energy
scale\cite{land}. Even if mini-black holes may be created at LHC,
probing the extra dimensions will require a very complicated and
delicate analysis of cornucopia of produced particles and their
interactions.

As far as extra dimensions are concerned, it would be interesting to
think about alternative methods to probe them. As an alternative, we
wondered how extra dimensions might influence, for instance, quantum
teleportation by using some newly developed technology. In this
paper, we investigate quantum entanglement and teleportation in a
more general situation -- in the spacetimes of higher dimensional
black holes. We will show how the extra dimensions and even
TeV-level gravity would change the properties of entanglement and
teleportation, when two observers stay on a 3-brane (our universe),
but not in the bulk. Our scheme differs from the standard
teleportation protocol in that one observer, Alice, stays stationary
at an asymptotically flat region of a black hole, while the other
observer, Bob, moves from the Alice's place toward the black hole.
We assume that both Alice and Bob hold an optical cavity for the
measurements of the inside state. They instantaneously share an
entangled Bell state at the same initial point in the flat region.
Then Bob picks his qubit and travels to the event horizon of the
black hole. Teleportation can be performed between Alice and Bob
when Bob stays at a fixed radius near the event horizon.

However, in contrast with the Minkowski spacetime, quantum
teleportation in a black hole spacetime requires several points to
be clarified such as Bob's trajectory and the shape of the cavity in
which particles are confined. It is well known that entanglement
states are sensitive to the environment and Bob's trajectory may
influence his description of the qubit state. For instance, if Bob's
cavity is not perfectly insulated, thermal particles from the black
hole might flow into the cavity. Fortunately, if Bob approaches the
black hole by freely falling, this situation can be avoided, since a
geodesic detector sees no particle coming out of the black hole
\cite{unruh}. Thus, after sharing the initial entangled state, Bob
may freely fall and then stop freely falling and become stationary
at a fixed radius, but to avoid the infinite deceleration due to an
instantaneous stop he may slowly decelerate for a while and then
become stationary at the fixed radius. During this process the
cavity might be teeming with thermal particles, which does not
matter seriously, because what is needed is the degree of
entanglement and teleportation under this condition.

On the other hand, if the cavity is totally reflecting and does not
couple at all to the outside, one can move Bob's cavity (a
rectangular box) adiabatically to near the black hole horizon, which
does not introduce any thermal particle of the Unruh vacuum inside
the cavity. The Boulware vacuum is the ground state for the inside
of a box that is stationary in a stationary black hole whose
timelike Killing vector is used to define the ground state, while
the Unruh vacuum is the Minkowski vacuum that has thermal particles
relative to the Boulware vacuum and can be detected by an
accelerated observer \cite{page}. Unruh and Wald discussed about
mining energy from a black hole by lowering an originally empty box
to a black hole sufficiently slowly, opening up the box and letting
the surrounding Unruh thermal radiation flow into the box
\cite{wald}. They also found that during this process, if one does
not open the box, then the interior of the box will remain in the
vacuum state with respect to its local time coordinate. The reason
is that the boundaries of the box that both move with an
acceleration effectively remove the horizon from consideration and
thereby keep the field inside the closed box in the vacuum state.

Of course, even for a well-insulated and reflecting cavity, one
cannot screen gravity and an initial vacuum state inside the cavity
might evolve into thermal Unruh states because of gravitational
coupling between the inside of the cavity and the outside. We
estimate a possible coupling time for a Boulware vacuum inside a
cavity, which perfectly reflects non-gravitational fields, to become
approximately thermal due to gravitational effects. For a thick wall
cavity outside a solar mass black hole, the coupling time is larger
than the age of our universe, which implies that the gravitational
coupling effect is negligible. Note that this works under the
condition that the size of  the cavity is much smaller compared to
the curvature of the black hole. Otherwise, the cavity cannot act as
it were in a thermal bath that is equilibrium with the black hole.
For a mini-black hole that may be created at LHC with typical
horizon curvature $10^{-19} \rm m$, it is hard to make a thick wall
box (say, of the size $1{\rm m}\times 1{\rm m}\times 1{\rm m}$) near
the horizon in thermal equilibrium. And also, the coupling between
the inside and the outside will diverge if one takes an infinitely
thin box and couples the inside with the outside across a tiny
distance.

Now our situation is that what we put into Bob's cavity initially is
not a pure state  but a state entangled with a particle inside
Alice's cavity. This entanglement will be maintained by any
evolution of the cavity as long as the cavity remains totally
reflecting and decoupled from any outside system. In this case, the
entanglement between Bob's particle and Alice's particle will
survive and perfect teleportation is still possible. This process
works well for teleportation outside massive black holes, but for
mini-black holes, especially those short-lived black holes that may
be created at LHC, it is impossible to lower Bob's cavity slowly to
approach black hole horizons.

The main purpose of this paper is to investigate possible influences
that extra dimensions might impose on the entanglement and
teleportation in curved spacetimes. So we do not discuss the
situation, in which both cavities evolve adiabatically, because this
makes no distinctions between teleportation in a curved spacetime
and that in a flat spacetime. Contrarily, we consider the situation,
in which Bob freely falls, then slow downs for a finite period of
time and stops freely falling and becomes stationary at a fixed
radius. Then, measurements of teleportation can be performed between
Alice and Bob. The results show that in a higher dimensional black
hole spacetime the degree of entanglement and the fidelity of
teleportation is closely related to the extra dimensions, the mode
frequency, the mass and the angular momentum per unit mass of the
back hole.

The organization of this paper is as follows. In Sec. II, we study
quantum entanglement and teleportation in a higher dimensional
Schwarzschild black hole spacetime. After reviewing a basic feature
of quantum field theory in this curved spacetime, we estimate the
gravitational coupling time for states inside and outside the cavity
and then go directly to the discussion of entanglement and
teleportation. In Sec. III, we extend our discussion to a rotating
black hole spacetime. The discussion applies not only to scalar
particles but also to gauge bosons and Dirac particles. Finally, we
discuss the effects of the extra dimensions, mode-frequency and
black hole parameters on fidelity of quantum teleportation in Sec.
IV.

\section{higher dimensional Schwarzschild black hole cases}
\subsection{Hawking radiation and vacuum structure}

We first review some essential feature of quantum field theory in a
higher dimensional Schwarzschild black hole, which is relevant to
quantum entanglement. It is the global structure of a black hole
spacetime that plays an important role in understanding Hawking
radiation and black hole thermodynamics \cite{hawking,unruh}. It has
also been known that the vacuum of a nonstationary curved spacetime
is in general not equivalent to the Minkowski vacuum. The event
horizon of a black hole causally disconnects the exterior region
from the interior one. The global spacetime structure of a uniformly
accelerated observer consists of two Rindler wedges: one is causally
disconnected from the other by the future and past horizons. Thus
the accelerated observer detects a thermal spectrum while moving
through the Minkowski vacuum, the so-called Unruh effect
\cite{unruh,takagi}. The spacetime near the event horizon of a
Schwarzschild black hole approximately looks like the Rindler
spacetime, which provides the Hawking radiation with an
interpretation of the Unruh effect. Observing that the causally
disconnected region surrounded by horizons provides a fictitious
system to an outside observer, Israel \cite{israel} applied
thermofield dynamics \cite{tu} to the Rindler wedge or black holes
to explain their thermal nature (see also \cite{vitiello}).

In the following, we use the two-photon state of the electromagnetic
field, which may be modeled by a massless scalar field, where the
polarization is ignored \cite{al2}. The metric of a $d$-dimensional
Schwarzschild spacetime is given by \cite{merys}
\begin{equation}
ds^2=-\left[1-\Bigl(\frac{r_{h}}{r} \Bigr)^{d-3}\right]dt^2
+\left[1- \Bigl(\frac{r_{h}}{r}
\Bigr)^{d-3}\right]^{-1}dr^2+r^2d\Omega^2_{d-2}, \label{schw}
\end{equation}
where $r_{h}$ denotes the event horizon with the area
$A_{d}=r^{d-2}_{h}\Omega_{d-2}$, where $\Omega_{d-2}$ is the volume
of a unit $(d-2)$-sphere. The mass of the $d$-dimensional black hole
is given by
\begin{equation}
\label{mbh} M=\frac{(d-2)r^{d-3}_{h}\Omega_{d-2}}{16\pi G_{d}}
\end{equation}
for the $d$-dimensional Newton's constant $G_{d}$. The Klein-Gordon
equation for the massless scalar field,
\begin{equation}
\label{klein}
\frac{1}{\sqrt{-g}}\partial_{\mu}(\sqrt{-g}g^{\mu\nu}\partial_{\nu}\phi)=0,
\end{equation}
has the positive frequency mode solution
\begin{equation}
\phi(t,r,\theta,\varphi) = e^{-i\omega t} \frac{R_{\omega
l}(r)}{r^{\frac{d-2}{2}}} Y_{lm}(\Omega),
\end{equation}
where $Y_{lm}(\Omega)$ is the $(3+n)$-spatial-dimensional
generalization of the usual spherical harmonic functions depending
on the angular coordinates, and $R_{\omega l}(r)$ satisfies the
radial equation
\begin{equation}
\label{radial} \frac{\partial^2 R_{\omega l}}{\partial
r_{*}^2}+\omega^{2}R_{\omega
l}-f(r)\left[\frac{(d-2)^2}{4r^2}f(r)+\frac{d-2}{2r}\frac{df}{dr}+\frac{l(l+n+1)}{r^2}\right]R_{\omega
l}=0.
\end{equation}
Here $n = d-4$ is the extra dimensions, and $f(r) =
[1-(r_{h}/r)^{d-3} ]$ and $r_{*}=\int dr/f(r)$ denotes a tortoise
coordinate, which may be exactly computed by expanding the
denominator in partial fractions to yield the result \cite{lawrence}
\begin{equation}
r_{*}=r+\frac{r_{h}}{d-3}\sum^{d-4}_{k=0}{\rm
ln}\left[\frac{r}{r_{h}}-e^{-i\frac{2\pi k}{d-3}}\right].
\end{equation}

To quantize the scalar field, we make use of the global structure of
the spacetime given by the Penrose diagram for the extended Kruskal
manifold in Fig. 1. The particle states are defined by positive
frequency modes or wave packets with respect to timelike Killing
vectors. Note that the metric (\ref{schw}) has a future-directed
Killing vector $\partial_t$ in $I$ and $\partial_{-t}$ in $II$ of
Fig. 1. Solving (\ref{radial}), we obtain the positive frequency
solutions
\begin{eqnarray}
\label{RI}
\phi_{I, p}=e^{-i\omega t}R_{\omega l}=e^{-i\omega (t+r_{*})}, \\
\label{RII} \phi_{II, p}=e^{-i\omega t}R_{\omega l}=e^{-i\omega
(t-r_{*})},
\end{eqnarray}
where $p$ stands for $(\omega, l, m)$.  The wave packets may be
found such that $\phi_{I, p}$ $(\phi_{II, p})$ have a support in $I$
$(II)$ but vanish in $II$ $(I)$, which may constitute a complete set
along a Cauchy surface $t = 0$ \cite{takagi}. Using (\ref{RI}) and
(\ref{RII}), the quantum field may be quantized as
\begin{equation}
\phi = \sum_{p}\left[b^I_{p}\phi_{I, p}+b^{II}_{p}\phi_{II, p
p}+b^{I \dagger}_{p}\phi^{*}_{I, p}+b^{II \dagger}_{p}\phi^{*}_{II,
p} \right]
\end{equation}
where the operators $b^{I}_{p}$ $(b^{II}_{p})$ and $b^{I
\dagger}_{p}$ $(b^{II \dagger}_{p})$ are the annihilation and
creation operators in region $I (II)$, respectively. The
Fulling-Rindler vacuum is defined as
\begin{eqnarray}
b^{I}_p \vert 0 \rangle_I \otimes \vert 0 \rangle_{II} = b^{II}_p
\vert 0 \rangle_I \otimes \vert 0 \rangle_{II} = 0.
\end{eqnarray}
Rewriting (\ref{RI}) and (\ref{RII}) in Kruskal coordinates as
\begin{eqnarray}
\label{u} \phi_{I} = (-U/a)^{i\omega a}, \quad \phi_{II} =
(U/a)^{-i\omega a},
\end{eqnarray}
where $a = 1/\kappa = 2r_{h}/(d-3)$ for the surface gravity
$\kappa$, (\ref{u}) may be analytically continued in the lower
half-plane of $U$ as
\begin{equation}
\label{anal} \tilde{\phi}_{II}=e^{\pi \omega a}(U/a)^{i\omega a}.
\end{equation}
\begin{figure}
\psfig{file=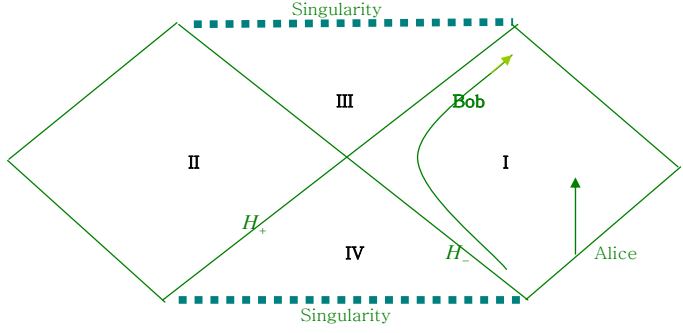,height=7.5in,width=5.5in }\caption{Penrose
diagram for the extended Kruskal manifold. }
\end{figure}

On the other hand, the Schwarzschild metric in Kruskal coordinates
takes the form
\begin{eqnarray}
\label{krs} &&ds^2 = - \left[1-\Bigl(\frac{r_{h}}{r}
\Bigr)^{d-3}\right]
dUdV+r^2d\Omega^2_{d-2},\nonumber\\
&&U = \mp \frac{2r_{h}}{d-3}e^{-\frac{d-3}{2r_{h}}u},\quad V = \pm \frac{2r_{h}}{d-3}e^{\frac{d-3}{2r_{h}}v},\nonumber\\
&&u =t-r_{*}, \quad v=t+r_{*},
\end{eqnarray}
where the upper (lower) sign is for $I (II)$. Now $\partial/\partial
U$ is another timelike Killing vector. The mode that is positive
(negative) with respect to $\partial/\partial U$ and extends over
the whole $V=0$, may be found
\begin{eqnarray}
\label{u2} \phi_{1} &=& \phi_{I}+ \tilde{\phi}_{II}=(-U/a)^{i\omega
a}+e^{\pi \omega a}(U/a)^{i\omega a},\\
\phi_{2} &=& \phi^{*}_{I}+\tilde{\phi}^{*}_{II}=(-U/a)^{-i\omega
a}+e^{-\pi \omega a}(U/a)^{-i\omega a}.
\end{eqnarray}
Note that the properly normalized mode $\phi_{1}/(e^{2\pi \omega
a}-1)^{1/2}$ leads to the spectrum of Hawking radiation. Using the
normalized modes
\begin{eqnarray}
\phi^{*}_{+} &=& e^{-\pi\omega a/2}\phi_{1}=e^{-\pi\omega
a/2}(-U/a)^{i\omega
a}+e^{\pi \omega a/2}(U/a)^{i\omega a},\nonumber\\
\phi^{*}_{-} &=& e^{\pi\omega a/2}\phi_{2}=e^{\pi\omega
a/2}(-U/a)^{-i\omega a}+e^{-\pi \omega a}(U/a)^{-i\omega a},
\end{eqnarray}
we quantize the scalar field as
\begin{equation}
\phi =\sum_{p} [2\sinh(\pi\omega
a)]^{-1/2}\left[d^{I}_{p}\phi_{+}+d^{II}_{p}\phi_{-}+d^{I\dagger}_{p}\phi^{*}_{+}+d^{II\dagger}_{p}\phi^{*}_{-}\right].
\end{equation}

The Kruskal operators are related with the Schwarzschild operators
by the Bogoliubov transformations
\begin{eqnarray}
b^{I}_{p} &=& [2\sinh(\pi\omega a)]^{-1/2}\left(e^{\pi\omega
a/2}d^{I}_{p}+e^{-\pi \omega
a/2}d^{II\dagger}_{p}\right),\nonumber\\
b^{II}_{p} &=& [2\sinh(\pi\omega a)]^{-1/2}\left(e^{\pi\omega
a/2}d^{II}_{p}+e^{-\pi \omega a/2}d^{I\dagger}_{p}\right).
\end{eqnarray}
The Bogoliubov transformation for each mode may be expressed through
a unitary transformation as
\begin{eqnarray}
b^{\sigma}_p = S_p (r) d^{\sigma}_p S^{\dagger}_p (t), \quad (\sigma
= I, II),
\end{eqnarray}
in terms of the two-mode squeeze operator
\begin{eqnarray}
S_p (r) &=& \exp [r (d^{I}_p d^{II}_p - d^{I \dagger}_p d^{II
\dagger}_p)] \nonumber\\
&=& \exp [- \tanh r (d^{I \dagger}_p d^{II \dagger}_p)] \exp[- \ln
\cosh r (d^{I \dagger}_p d^{I}_p + d^{II \dagger}_p d^{II}_p +1)]
\exp [\tanh r (d^{I}_p d^{II}_p)],
\end{eqnarray}
where $\tanh r = e^{-\pi \omega a}$. The inverse transformation is
given by
\begin{eqnarray}
d^{\sigma}_p = T_p (r) b^{\sigma}_p T^{\dagger}_p (t), \quad (\sigma
= I, II),
\end{eqnarray}
where $T_p (r)$ is another two-mode squeeze operator
\begin{eqnarray}
T_p (r) &=& \exp [- r (b^{I}_p b^{II}_p - b^{I \dagger}_p b^{II
\dagger}_p)] \nonumber\\
&=& \exp [\tanh r (b^{I \dagger}_p b^{II \dagger}_p)] \exp[- \ln
\cosh r (b^{I \dagger}_p b^{I}_p + b^{II \dagger}_p b^{II}_p +1)]
\exp [- \tanh r (b^{I}_p b^{II}_p)].
\end{eqnarray}

Now the Minkowski vacuum defined by
\begin{eqnarray}
d^{I}_p \vert 0 \rangle_M = d^{II}_p \vert 0 \rangle_M = 0,
\end{eqnarray}
is simply given by
\begin{eqnarray}
\vert 0 \rangle_M &=& \prod_{p} T_p (r) \vert 0 \rangle_I \otimes
\vert 0 \rangle_{II}\nonumber\\ &=& \prod_{p} \cosh^{-1} r \exp
[\tanh r (b^{I \dagger}_p b^{II
\dagger}_p) ] \vert 0 \rangle_I \otimes \vert 0 \rangle_{II} \nonumber\\
&=&\sum_{n_{p}=0}^{\infty}\prod_{p} \cosh^{-1}r \tanh^{n_{p}} r
\vert n_{p} \rangle_{I} \otimes
 \vert n_{p} \rangle_{II}. \label{vac}
\end{eqnarray}
where $\vert n_{p} \rangle_{I} = (b^{I \dagger}_p)^n \vert 0
\rangle_I /\sqrt{n!}$ and $\vert n_{p} \rangle_{II} = (b^{II
\dagger}_p)^n \vert 0 \rangle_{II} /\sqrt{n!}$ are orthonormal bases
for Hilbert space $H_{I}$ and $H_{II}$, respectively. Equation
 (\ref{vac}) shows that the original vacuum state evolves into an
 Einstein-Podolsky-Rosen type correlation. Also note that the vacuum (\ref{vac})
 is an extension to $I$ and $II$ of the thermal state
 in $I$ with the inverse temperature $\beta = 2 \pi a = 2 \pi /\kappa$, which
 is the essence of thermofield dynamics \cite{tu}. Thus the vacuum (\ref{vac})
 and its excited states are mixed states of
 outgoing particles in $H_{I}$. For an observer outside the
 black hole, quantum unitarity is lost because he would not be able to
 do any measurement in $H_{II}$. As $H_{II}$ is no longer accessible to
 him, he can only make an
 average over the states in $H_{II}$ to obtain the density operator
 in $H_{I}$. In fact, the measurement of an operator  ${\cal O}_I$ in $I$
 is the trace of ${\cal O}_I$ weighted with the thermal operator
 $\rho_I = e^{ - \beta b^{I \dagger} b^I}/Z_I$, ($Z_I = Tr \rho_I$).

One can simplify the analysis by considering the effect of
teleportation of the target state
 $\vert \varphi \rangle_{M} = a \vert 0 \rangle_{M} + b \vert 1 \rangle_{M}$
 by the Minkowski observer Alice to a single Schwarzschild mode of the
 observer Bob. Thus, one can only consider the mode $p$ in region I which is distinct
 from the negative  mode in the same region \cite{al2}. We emphasize
 that both Alice's and Bob's cavities are designed to detect
 particles, so we neglect the antiparticle modes in their cavities in the hereafter
 discussions.
 Therefore, the single-mode component of the Minkowski vacuum
state, namely the two-mode squeezed state, is given by
\begin{eqnarray}
\label{zero}
 \vert 0 \rangle_{M}^{\mathcal{~B}} = \frac{1}{\cosh r}\sum^{\infty}_{n=0}
\tanh^{n}r \vert n \rangle_{I} \otimes \vert n \rangle_{II},
\end{eqnarray}
and, similarly, the excited state by
\begin{eqnarray}
\label{one} \vert 1 \rangle_{M} ^{\mathcal{~B}}&=& d^{I \dagger}_p
\vert 0
\rangle_{M} \nonumber\\
&=& \frac{1}{\cosh ^2 r}\sum^{\infty}_{n=0} \tanh^{n}r \sqrt{n+1}
\vert n+1 \rangle_{I} \otimes \vert n\rangle_{II}.
\end{eqnarray}

\subsection{Entanglement and teleportation}

We now discuss how Alice and Bob come to share an entangled resource
for teleportation. After the coincidence of Alice and Bob, Alice
remains at the asymptotical flat region, while Bob slowly lowers
himself and his cavity to the vicinity of the black hole horizon.
 As noted previously, we assume the inside of Bob's cavity is
 totally decoupled from the outside. Thus, when the cavity is
 lowered to a fixed radius outside the horizon and becomes
 stationary, the entanglement between Bob's particle and Alice's
 particle will be exactly maintained. In this case, perfect
 entanglement and teleportation is possible and for massive black holes, this protocol works well.
 
However, a puzzle might be raised that nobody can screen gravity, so
there would be some gravitational coupling between the inside and
the outside of the cavity, even if the cavity is well-insulated and
reflecting all non-gravitational fields. Hence, the Boulware vacuum
inside the
 cavity that is perfectly insulated from non-gravitational fields
 from the outside, might couple gravitationally to the outside and evolve into
 approximately the Unruh thermal state. We estimate the coupling time for the inside to become approximately
 thermal through the gravitational interaction. Let us consider a
 thick-wall box (for example, the thickness of the bottom wall is
  $l=1 {\rm m}$), which is dropped very near the horizon of a Schwarzschild black hole,
  and the mass of the black hole is assumed to be of order of
  a solar mass ($M_{bh}\sim M_{\odot}$) and thus the black hole
  temperature is about $T_{bh}\sim 10^{-8}~K$. The size of the cavity
  must be much smaller than that of a black hole, otherwise
  it cannot maintain thermal equilibrium with the Hawking radiation.

Now no fields are confined to
  this rectangular box with an inside volume $V_{c}=1{\rm m} \times 1{\rm m} \times 1{\rm m}$.
  Then, the cavity outside the horizon will respond as though it
  were in thermal bath of temperature $T=T_{bh}/\chi$, where $\chi$ is
  the redshift factor defined by $\chi=\sqrt{-g_{00}}$. The energy
  density of thermal radiation is given by the Stefan-Boltzmann
  law
  \begin{equation}
  \label{boltz}
\rho=\alpha T^{4}, ~~~\alpha=\frac{\pi^2 k_B^4}{15 \hbar^3 c^3}
  \end{equation}
  We assume that after a time interval $t$, the box will be teeming
  with thermal flux.
Hence, from an infinity observer viewpoint,  a graviton with energy
$h\nu_{0}$, crossing the thick walls of the cavity obey the
following equation
\begin{equation}
  \label{graviton}
h\nu\approx h\nu_{0}-\frac{h\nu_{0}}{c^2}\kappa \Delta l,
  \end{equation}
  where $\kappa = 2\pi ck_{B}T_{bh}/\hbar$ is the surface gravity of the black
  hole and $\Delta
l=1 m$. We then have
\begin{equation}
  \label{graviton}
\frac{\Delta \nu}{\nu_{0}}=\frac{\kappa}{c^2}\Delta l.
  \end{equation}
A clock inside the cavity will become slower since the cavity filled
with a thermal radiation  is heavier than before. Thus the change of
the clock running from Eq. (\ref{graviton}) is
\begin{equation}
  \label{deltatime}
\frac{\Delta t}{t}=\frac{\kappa}{c^2}\Delta l.
  \end{equation}
  According to the uncertainty principle, the uncertainty of time
  is attributed to the change of energy inside the cavity, that is,
 $\Delta E \Delta t\sim h$, where $\Delta E=\alpha
  T_{bh}^{4}V_{c}$. Finally, we find that
\begin{equation}
  \label{coulptime}
{t}\sim \frac{h}{\frac{{\kappa}}{c^2}\Delta l\Delta E}.
  \end{equation}
  The results show that in our case $t\sim 10^{19}
  s$, which is longer than the present age of our universe ($\approx 4\times 10^{17}s$).
  Therefore, the gravitational coupling between the inside and the outside
 of the cavity is negligible. For mini-black holes, the above
  estimation about gravitational coupling time does not work in that
  it is difficult to define a thick-wall cavity surrounded by a thermal bath near the black hole horizon. Therefore,
  for massive black holes, Bob can slowly lower his cavity down to
  the black hole horizon and the state inside the cavity will evolve
  adiabatically. Since gravitation coupling is negligible, perfect entanglement and
 teleportation is still possible for observers with a well insulated cavity outside
 a black hole. However, for mini-black holes, it is technically difficult to
 consider a very thin cavity to perform teleportation
 without losing information.

 As emphasized in Sec.~I, in this paper, we focus on
 the condition that Bob can freely fall toward the black hole and then
stop through a slow acceleration on a surface outside the event
horizon, and Bob's cavity evolves not totally adiabatically and
becomes full of thermal radiation near the
 black hole horizon.
  Then the
 state inside Bob's cavity is no longer perfect entangled with that
 of Alice.
 One can assume that prior to their coincidence,
 Alice and Bob have no photons in
their cavities. Suppose that each cavity supports two orthogonal
modes, with the same frequency, labeled $A_{i}$ and $B_{i}$ with
$i=1,2$, which are each excited to a single photon Fock state at the
coincidence point. The state held by Alice and Bob is then the
entangled Bell state
\begin{equation}
\label{bell} \vert \phi \rangle_{M} = \frac{1}{\sqrt{2}}\left(\vert
0 \rangle_{M}^{\mathcal{~A}} \vert
    0 \rangle_{M}^{\mathcal{~B}}
    +\vert 1 \rangle_{M}^{\mathcal{~A}}\vert 1 \rangle_{M}^{\mathcal{~B}}\right),
    \end{equation}
where the first qubit in each term refers to Alice's cavity with
index $\mathcal{~A}$, the second qubit to Bob's cavity with index $
\mathcal{~B}$. The Bell state is a maximally entangled state in
inertial frames. If Bob undergoes a uniform acceleration or stays in
curved space-time, the state in his cavity must be specified in
Rindler or Schwarzschild coordinates. As a consequence, the second
state in each term of (\ref{bell}) should have a Schwarzschild mode
expansion given by (\ref{zero}) and (\ref{one}). We can then rewrite
Eq. (\ref{bell}) in terms of Minkowski modes for Alice and
Schwarzschild modes for Bob. Since Bob is causally disconnected from
region $II$, we must trace over the states in this region, which
results in a mixed
     state \cite{mann}
     \begin{eqnarray}
&&\rho_{AB}=\frac{1}{2 \cosh^2r}\sum_{n}(
\tanh r)^{2n}\rho_{n}, \nonumber\\
&&\rho_{n} = \vert 0 \rangle \langle 0,n \vert
+\frac{\sqrt{n+1}}{\cosh r} \vert 0,n \rangle \langle 1,n+1 \vert
+\frac{\sqrt{n+1}}{\cosh r} \vert 1,n+1 \rangle
 \langle 0,n \vert + \frac{{n+1}}{\rm cosh^2 r}\vert 1,n+1 \rangle
 \langle 1,n+1 \vert,
 \end{eqnarray}
 where $\vert n,m \rangle = \vert n \rangle_{M}^{\mathcal{~A}} \vert m \rangle_{I}^\mathcal{~B}$.
 The partial transpose of the density operator $\rho_{AB}$ can be
 obtained by interchanging Alice's qubits as
\[
\rho_{AB}^{T}=\frac{\tanh^{2n}r}{2 \cosh^2r} \left(
\begin{array}{cccc}
1 & 0 & 0 & 0\\ 0 & 0 & \frac{\sqrt{n+1}}{\cosh r} & 0
\\0 & \frac{\sqrt{n+1}}{\cosh r}  & 0 & 0\\0 & 0& 0 & \frac{{n+1}}{\cosh r}
\end{array}
\right),
\]
and the corresponding negative
     eigenvalue of the partial transpose is given by
\begin{equation}
\lambda_{n}=-\frac{\tanh^{2n}r}{2 \cosh^3r}\sqrt{n+1} ,
    \end{equation}
   The degree of entanglement for the two observers here can be
 measured by using the the concept of logarithmic negativity
 \cite{vidal}. The logarithmic negativity is defined as
\begin{equation}
E_{\mathcal{N}}(\rho)\equiv  \log_{2}(2\mathcal{N}(\rho)+1),
    \end{equation}
    where $\mathcal{N}(\rho)$ is the negativity of
    the state. The negativity is defined as the absolute sum of the
    negative eigenvalues of the partial transpose with respect to $\rho_{\rm
    AB}^{T}$. So
\begin{equation}
\mathcal{N}(\rho)\equiv  \sum_{n}
\frac{|\lambda_{n}|-\lambda_{n}}{2}.
\end{equation}
The entanglement monotone is found to be
\begin{equation}
\label{negativity} E_{\mathcal{N}}(\rho)=
\log_{2}\left(1+\sum_{n=0}^{\infty}\frac{\tanh^{2n}r}{
\cosh^3r}\sqrt{n+1}\right).
\end{equation}
At $r\rightarrow 0$, corresponding to $\kappa\rightarrow 0$, we are
back to the usual case of entanglement between Alice and Bob where
both are Minkowski  observers, and there are no Schwarzschild modes
in Bob's cavity (i.e. $n=0$). And thus $E_{\mathcal{N}}(\rho)=1$.
For a finite acceleration the entanglement is degraded. In the limit
$r\rightarrow \infty$, the second term in (\ref{negativity}) is
vanishing. The logarithmic negativity is then $0$. Figure 2 shows
that for a finite surface gravity, the degree of entanglement is
reduced. Figure 3 shows that if the black hole mass $M$ is fixed,
the logarithmic negativity is reduced as the extra dimension number
$n$ increases. On the other hand, if $n$ is fixed, $E_{\mathcal{N}}$
is enhanced as $M$ increases.
\begin{figure}
\psfig{file=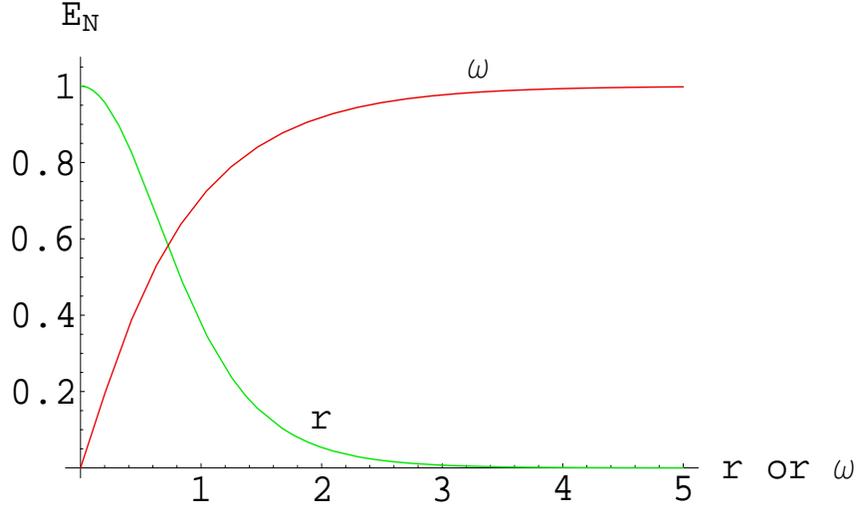, height=3in, width=4.5in }\caption{The
logarithmic negativity as a function of $r$ $(e^{2r} = {\rm coth}
(\pi \omega/ 2\kappa))$, and the mode frequency $\omega$. The green
line shows that the entanglement monotone reduces with $r$ when
$\omega$ is fixed, while the red line shows that for the fixed
surface gravity the entanglement monotone increases with $\omega$. }
\end{figure}
\begin{figure}
\psfig{file=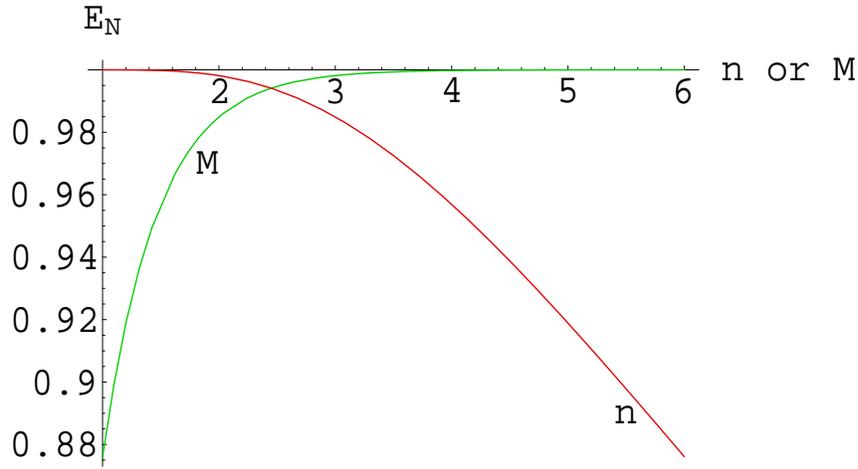, height=3in, width=4.5in }\caption{The
logarithmic negativity as a function of the extra dimensions
($n=d-4$) and the black hole mass $M$.}
\end{figure}
\begin{figure}
\psfig{file=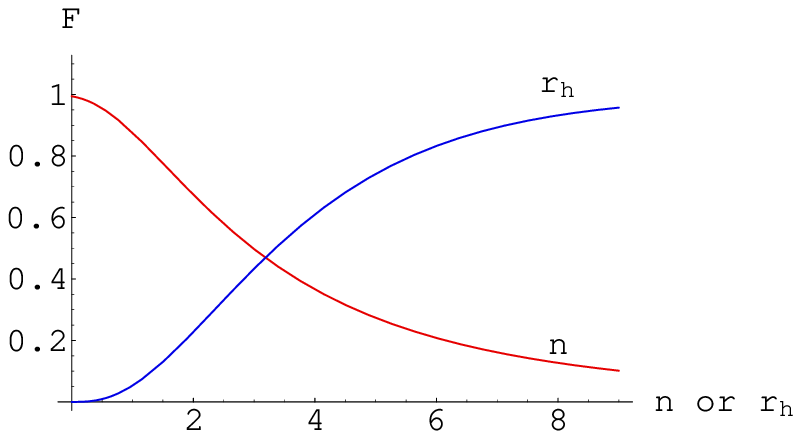, height=3in, width=4.5in }\caption{The
fidelity as the function of the extra dimensions $n$ and horizon
radius $r_{h}$. The fidelity decreases with the extra dimensions,
while the fidelity increases with the horizon radius. }
\end{figure}

     The states $\vert 0 \rangle_{M}$, $\vert 1\rangle_{M}$ are defined
     in terms of the physical Fock states for Alice's cavity by the dual-rail basis states as
     suggested by Ref. \cite{al2}: $\vert 0 \rangle_{M}=\vert 1
     \rangle_{A_{1}}\vert 0\rangle_{A_{2}}$, $\vert
     1\rangle_{M}=\vert
     0\rangle_{A_{1}}\vert 1 \rangle_{A_{2}}$, and the similar expressions for
     Bob's cavity. In order to teleport the unknown state
     $\vert \varphi \rangle_{M} = \alpha \vert 0\rangle_{M}
     +\beta \vert 1\rangle_{M}$
     to Bob, we should assume that Alice has an additional cavity,
     which contains a single qubit with dual-rail encoding by a
     photon excitation of a two-mode incoming (Minkowski) vacuum state. This
     will allow Alice to perform a joint measurement on the two
     orthogonal modes of each cavity. After Alice's measurement,
     Bob's photon will be projected according to the
     measurement outcome. The final state Bob received can be given by
      $\vert \varphi_{ij} \rangle =x_{ij}\vert 0 \rangle +y_{ij}\vert 1 \rangle$, where there
      are four possible conditional state amplitudes expressed as
      $(x_{00},y_{00})=(\alpha,\beta),
      (x_{01},y_{01})=(\beta,\alpha),(x_{10},y_{10})=(\alpha,-\beta)$, and
      $(x_{11},y_{11})=(-\beta,\alpha)$. Once receiving Alice's results of
     measurement, Bob can apply a unitary transformation to verify
     the protocol in his local frame. However, Bob must be confronted
     with   the fact that his cavity will become teemed with thermally excited
     photons because of the Hawking effect.

 When Alice sends the result of her measurement to
Bob, the state he observes must be traced out over region $II$,
since Rob is causally disconnected from region $II$ \cite{al2},
\begin{eqnarray}
&& \rho^{(I)}_{ij}=Tr_{II} (\vert \varphi_{ij} \rangle_{M} \langle
\varphi_{ij} \vert)=\sum^{\infty}_{n=0}p_{n}\rho^{I}_{ij,n}
 \nonumber\\
 &&=\frac{1}{\cosh^{6}r}\sum^{\infty}_{n=0}\sum^{n}_{m=0}\left[(\tanh^{2}r)^{n-1}[(n-m)|x_{ij}|^{2}+m
 |y_{ij}|^{2}]\right.\nonumber\\
 &&\left.\times \vert m,n-m \rangle_{I} \langle m,n-m \vert
 +(x_{ij}y^{*}_{ij} \tanh^{2n}r\right
 . \nonumber\\
 &&\left .\sqrt{(m+1)(n-m+1)})\times \vert m,n-m+1\rangle_{I}\right
 . \nonumber\\
 &&\left .\langle m+1,n-m \vert + {\rm H.c.})\right],
 \end{eqnarray}
  in particular with
 \begin{equation}
   \quad p_{0}=0, \quad p_{1}=1/\cosh^{6}r, \quad p_{n}=\frac{(\tanh^{2}r)^{n-1}}{\cosh^{6}r} .
  \end{equation}
  Since what we are concerned is to which extent $\vert \varphi_{ij} \rangle$
  might deviate from unitarity, it is reasonable for us to
  perform a unitary transformation on $\vert \varphi_{ij} \rangle$
  and convert its form into
  $\vert \varphi \rangle_{I}$. Suppose upon receiving the result $(i,j)$
  of Alice's measurement, Bob can apply the rotation operators
  restricted to the 1-excitation sector of his state.
  In this way, we may define the fidelity  corresponding to
  the teleportation as
\begin{eqnarray}
\label{fidelity}
  &&F^{I}(\vert \varphi \rangle)={}_{I} \langle \varphi \vert
  \rho^{I}\hat{U}\vert \varphi_{ij} \rangle_{I}
  \nonumber\\&&={}_{I} \langle \varphi \vert \rho^{I}\vert
  \varphi \rangle_{I}=(1-e^{-\pi
\omega/\kappa})^{3},
\end{eqnarray}
which is identical with the results of
  Alsing and Milburn.  From Fig. 4, we can see that the fidelity of teleportation
  depends on the extra dimension number and horizon radius.

We notice that the Hawking radiation emitted by larger four
   dimensional, astrophysical black holes has not yet been observed
   because for a black hole with several solar mass $M_{\odot}$, the
   characteristic temperature is about $T_{H}\sim 10^{-8}$K: an extremely low
    temperature corresponding to a very low energy frequency that
    cannot be detected. The primordial black holes
    that would have been created at the
    early universe with possible mass ($M_{BH}\sim 10^{15}$gr) and
the corresponding Hawking spectrum with a peak in the range $10-100~
{\rm
    Mev}$ have not been detected either.

In recent years, it is proposed that in brane-world theories the
    true scale $M_{*}$ of quantum gravity may be lower than the
    traditional Planck scale $M_{pl}$, possibly approaching TeV
    scales \cite{arkani}. The observed weakness of gravity at long distances is due to
    the presence of $n$ new spatial dimensions large compared to the electroweak scale. This follows
    that $M^{2}_{pl}\sim R^{n}M^{n+2}_{*}$, where $R$ is the size of the extra dimensions.
    A black hole with horizon radius $r_{h}$ smaller than
    the size of extra dimensions $R$, would submerge into extra
    dimensions, and for a given mass $M$ such a black hole
    becomes lighter, larger and colder than a usual four dimensional
    black hole with the same mass. If we assume that $M_{*}=1~{\rm
    TeV}$, which implies that $R\sim 10^{30/n - 17}{\rm cm}$
    and, choose the black hole mass $M_{BH}=5~{\rm
    TeV}$, the ratio between the $(4+n)$-dimensional Schwarzschild
    horizon radius and the ordinary 4-dimensional Schwarzschild with the
    same mass $M_{BH}=5~{\rm TeV}$ is given by \cite{arg}
\begin{equation}
  \frac{r_{h(4)}}{r_{h(4+n)}}\sim
  \left(\frac{r_{h(4+n)}}{R}\right)^n.
\end{equation}For $n=2$,
we have $r_{h(4+n)}=2.6 \times 10^{-19}{~\rm m}$ and then
$r_{h(4)}/r_{h(4+n)} \sim 10^{-30}$. So for a given mode frequency
$\omega$, the TeV-level gravity enhances the entanglement monotone
and teleportation fidelity greatly. But for a mini-black hole that
may be created at LHC with emitted Hawking radiation frequency range
$0.05 \lesssim \omega r_{h} \lesssim 0.5$, the logarithmic
negativity for entanglement is then bounded to $E_{\mathcal{N}\rm
max}=0.77$ and the teleportation saturates to $F_{\rm max}=0.25$.

\section{higher dimensional rotating black hole cases}

In this section, we discuss entanglement and teleportation in the
background spacetime of a higher dimensional rotating black hole.
The spacetime around a $(4+n)$-dimensional, rotating, uncharged
black hole is given by the following line-element \cite{merys}
\begin{equation}
  ds^2=\left(1-\frac{\mu}{\Sigma r^{n-1}}\right)dt^2+\frac{2a\mu \sin^2 \theta}
  {\Sigma r^{n-1}}dtd \varphi-\frac{\Sigma}{\Delta}dr^2-\Sigma
  d\theta^2
  -\left(r^2+a^2+\frac{a^2 \mu \sin^2 \theta}{\Sigma r^{n-1}}\right)
  \sin^2 \theta d \varphi^2-r^2 \cos^2 \theta d\Omega_{n},
\end{equation}
where
\begin{equation}
\Delta=r^2+a^2-\frac{\mu}{r^{n-1}}, \quad \Sigma = r^2 + a^2 \cos^2
\theta,
\end{equation} and $d\Omega_{n}$ is the line-element on a unit
$n$-sphere. The mass and the angular momentum per unit mass of the
black hole are given by,
\begin{equation}
M=\frac{(n+2)A_{n+2}}{16\pi G_{4+n}}\mu, \quad \frac{J}{M}
=\frac{2}{n+2}a,
\end{equation}with $G_{4+n}$ being the $(4+n)$-dimensional Newton's
constant, and $A_{n+2}$ the area of a $(n+2)$-dimensional unit
sphere given by
\begin{equation}
A_{n+2}=\frac{2\pi^{(n+3)/2}}{\Gamma[(n+3)/2]}.
\end{equation}
The horizon occurs when $\Delta(r)=0$, i.e., when $r=r_{h}$ with
\begin{equation}
r_{h}=\left[\frac{\mu}{1+a_{*}}\right]^{1/(n+1)},
\end{equation}
where $a_{*} = a / r_{h}$. Note that there is only a single horizon
when $n\geq 1$, in contrast to the four-dimensional Kerr black hole,
which has an inner and an outer horizon. We consider $\mu$ and $a$
as the normalized mass and angular momentum parameters,
respectively. Also note that there is no upper bound on $a$ when
$n\geq 2$, in contrast to the four dimensional case when $a$ is
bounded. The surface gravity and the angular velocity at the horizon
are given by \cite{park}
\begin{equation}
\kappa=\frac{(n+1)+(n-1)a^2_{*}}{2(1+a^2_{*})r_{h}}, \quad
\Omega=\frac{a_{*}}{(1+a^2_{*})r_{h}}.
\end{equation}
Greybody factors for five-dimensional rotating black holes on the
brane were calculated in Ref. \cite{park}.

In this section, we will investigate quantum entanglement and
teleportation not only with scalar particles, but also with gauge
bosons and Dirac particles in the spacetime of a higher-dimensional
rotating black hole. The angular momentum parameter $a$ (or $a_{*}$)
will definitely affect the degree of entanglement and fidelity of
teleportation. In order to derive a master equation describing the
motion of a field with arbitrary spin $s$, we need to make use of
the Newman-Penrose formalism \cite{penrose,chan}. The master
equations for a spin $s$ field, under the standard decomposition
\begin{equation}
\phi=R_{s}(r)S(\theta)e^{-i\omega t+im\varphi}, \label{kerr sol}
\end{equation}
read as \cite{park,kanti}
\begin{eqnarray}
\label{master}
&&\Delta^{-s}\frac{d}{dr}\left(\Delta^{s+1}\frac{dR_{s}}{dr}\right)+\left[\frac{K^2-i[2r+(n-1)\mu
r^{-n}]K}{\Delta}+4is\omega
r+s(\Delta''-2)-\Lambda_{sj}\right]R_{s}=0,\\
&&\frac{1}{\sin\theta}\frac{d}{d
\sin\theta}\left(\sin\theta\frac{dS}{d\theta}\right)+\left[-\frac{2m
s \cot\theta}{\sin\theta}-\frac{m^2}{\sin^2\theta}+a^2\omega^2
\cos^2 \theta-2as\omega
\cos\theta+s-s^2\cot^2\theta+\lambda_{sj}\right]S=0,
\end{eqnarray}
where
\begin{eqnarray}
K=(r^2+a^2)\omega-am, \quad \Lambda_{sj}=\lambda_{sj}+a^2\omega^2-2am\omega,\nonumber\\
{\rm and} \quad
\lambda_{sj}=j(j+1)-s(s+1)-\frac{2ms^2}{j(j+1)}a\omega + \cdots.
\end{eqnarray}
The asymptotic solutions of Eq. (\ref{kerr sol}) coming out the
horizon are given by
\begin{eqnarray}
\label{four}\phi_{I}=e^{-i(\omega-\omega_{0})\left( \hat{t}-r_{*}\right)},\\
 \label{five}\phi_{II}=e^{-i(\omega-\omega_{0})\left( r_{*}-\hat{t}\right )},
\end{eqnarray}
where
\begin{equation}
\frac{dr_{*}}{dr}=\frac{r^2+a^2}{\Delta}, \quad \omega_{0}=m\Omega,
\quad {\rm and} \quad \hat{t}=\frac{\omega}{\omega-\omega_{0}}t
\end{equation} where $r_{*}$ is the tortoise coordinate. By defining
the generalized light-like Kruskal coordinates \cite{dumo}
\begin{eqnarray}
&&U =-\frac{1}{\kappa}e^{-\kappa u},\quad
V=\frac{1}{\kappa}e^{\kappa v},
\quad {\rm for ~~~r>r_{h}},\\
&& U=\frac{1}{\kappa}e^{-\kappa u},\quad V=\frac{1}{\kappa}e^{\kappa
v}, \quad {\rm for ~~~r<r_{h}},\\
&&u=\hat{t}-r_{*}, \quad v=\hat{t}+r_{*}
\end{eqnarray}
we can rewrite (\ref{four}) and (\ref{five}) in the following form,
\begin{eqnarray}
\label{kfour}\phi_{I}={\rm exp}[\frac{i(\omega-\omega_{0})}{\kappa}{\rm ln}(-\kappa U)],\\
 \label{kfive}\phi_{II}={\rm exp}[\frac{i(\omega-\omega_{0})}{\kappa}{\rm ln}(\kappa
 U)].
\end{eqnarray}
By using the formula $-1=e^{i\pi}$ and making (\ref{kfour}) analytic
in the lower half-plane of $U$, we find a complete basis for
positive energy $U$ modes
\begin{eqnarray}
\label{kour}
&&\phi_{1}=(-U/a)^{\frac{i(\omega-\omega_{0})}{\kappa}}+e^{\frac{\pi(\omega-\omega_{0})}{\kappa}}
(U/a)^{\frac{i(\omega-\omega_{0})}{\kappa}},\\
 \label{kive}
 &&\phi_{2}=(-U/a)^{\frac{-i(\omega-\omega_{0})}{\kappa}}+e^{\frac{-\pi(\omega-\omega_{0})}{\kappa}}
(U/a)^{\frac{-i(\omega-\omega_{0})}{\kappa}}.
\end{eqnarray}
The Hawing radiation spectrum can be obtained by using
$(N_{\omega}\phi_{1},N_{\omega}\phi_{1})=N^2_{\omega}(1\pm
e^{\frac{2\pi(\omega-\omega_{0})}{\kappa}})=-1$, i.e.,
\begin{equation}
N^2_{\omega}=\frac{1}{e^{\frac{2\pi(\omega-\omega_{0})}{\kappa}}\pm
1},
\end{equation}
where $+$ corresponds to fermions and $-$ bosons. Similarly to the
Schwarzschild case, the vacuum in scalar and Maxwell fields can also
be written in the two-mode squeezed state,
\begin{equation}
\label{zer}
 \vert 0 \rangle_{M}= \frac{1}{\cosh r}\sum^{\infty}_{n=0}
\tanh^{n} r \vert n \rangle_{I} \otimes \vert n \rangle_{II},
\end{equation}
where $\cosh r=(1-e^{-2\pi (\omega-\omega_{0})/\kappa })^{-1/2}$ and
$\tanh r = e^{\pi (\omega-\omega_{0})/\kappa}$.

For Dirac fields, the Bogoliubov transformations between the Kruskal
and Schwarzschild operators are given by \cite{Iyer}
\begin{eqnarray}
a^{I}_{p}=\cos r c^{I}_{p}+\sin r c^{II \dagger}_p,\nonumber\\
a^{II}_{p} =\cos r c^{II}_{p} + \sin r c^{I \dagger}_p,
\end{eqnarray}
where the fermionic Bogoliubov coefficients allow us to define
\begin{equation}
\cos r = (1+e^{-2\pi(\omega-\omega_{0})/\kappa})^{-1/2}, \quad \sin
r =e^{-\pi(\omega-\omega_{0})/\kappa}(1+e^{-2\pi(\omega-\omega_{0})
/\kappa})^{-1/2}.
\end{equation}
As for the scalar field case, the Bogoliubov transformation can also
be expressed as
\begin{eqnarray}
a^{\sigma \dagger}_{p} = \tilde{S}_p (r) c^{\sigma \dagger}_{p}
\tilde{S}^{\dagger}_p (r), \quad c^{\sigma \dagger}_{p} =
\tilde{T}_p (r) a^{\sigma \dagger}_{p} \tilde{T}^{\dagger}_p (r),
\quad (\sigma = I, II)
\end{eqnarray}
in terms of the two-mode squeeze operators
\begin{eqnarray}
\tilde{S}_p (r) = \exp [- r (c^{I}_{p} c^{II}_{p} - c^{I
\dagger}_{p}c^{II \dagger}_{p})], \quad \tilde{T}_p (r) = \exp [r
(a^{I}_{p} a^{II}_{p} - a^{I \dagger}_{p} a^{II \dagger}_{p})]
\end{eqnarray}
Then the Dirac vacuum defined by $ a^{\sigma}_{p}\vert 0 \rangle_M =
0$ is given by
\begin{eqnarray}
\label{dirac} \vert 0 \rangle_M &=& \tilde{T}_p (r) \vert 0
\rangle_I
\otimes \vert 0 \rangle_{II} \nonumber\\
&=& e^{\ln \cos r }\exp[ \ln (1 +  \tan r c_{p}^{I \dagger}
c_{\tilde{p}}^{II\dagger})] \vert 0 \rangle_{I} \otimes \vert 0
\rangle_{II}.
\end{eqnarray}
Here $N_{f}=\cos r$ is the fermionic normalization factor.
Considering the finite number of allowed excitations in the
fermionic system due to the Pauli exclusion principle, Eq.
(\ref{dirac}) can be rewritten as
\begin{eqnarray}
\label{dirac2}
 \vert 0 \rangle_{M} = N_{f}\{\vert 0\rangle_{I} \otimes \vert 0\rangle_{II}+
\sum_{p} \tan r\left(\vert 1_{p}\rangle_{I}\right. \left. \otimes
\vert 1_{\tilde{p}}\rangle_{II}+\vert
1_{\tilde{p}}\rangle_{I}\otimes \vert 1_{p}\rangle_{II}\right)\}.
\end{eqnarray}
Since we only need to consider the information teleported out of the
black hole, we can drop the second set of parentheses in Eq.
(\ref{dirac}). We now simplify our analysis by considering the
effect of teleportation of the initial quantum state inside the
horizon to a single mode of the outside Hawking particles, which
goes as
\begin{equation}
\vert 0 \rangle_{M}= \cos r \vert 0 \rangle_{I} \otimes \vert
0\rangle_{II}+ \sin r \vert 1\rangle_{I} \otimes \vert
1\rangle_{II}.
\end{equation}

\subsection{Quantum entanglement and
teleportation with scalar particles and gauge bosons}

In this subsection, we discuss quantum entanglement and
teleportation with scalar particles and gauge bosons in the rotating
black hole spacetime. The form formula for entanglement monotone is
the same as that of the Schwarzschild case, except for the
difference that now the angular momentum parameter contributes to
the value of the surface gravity. Thus, the logarithmic negativity
as a function of the angular momentum parameter $a$ (or $a_{*})$ is
written as
\begin{equation}
\label{bnegativity} E_{\mathcal{N}}(\rho)=
\log_{2}\left(1+\sum_{n=0}^{\infty}\frac{\tanh^{2n}r}{
\cosh^3r}\sqrt{n+1}\right).
\end{equation}
\begin{figure}
\psfig{file=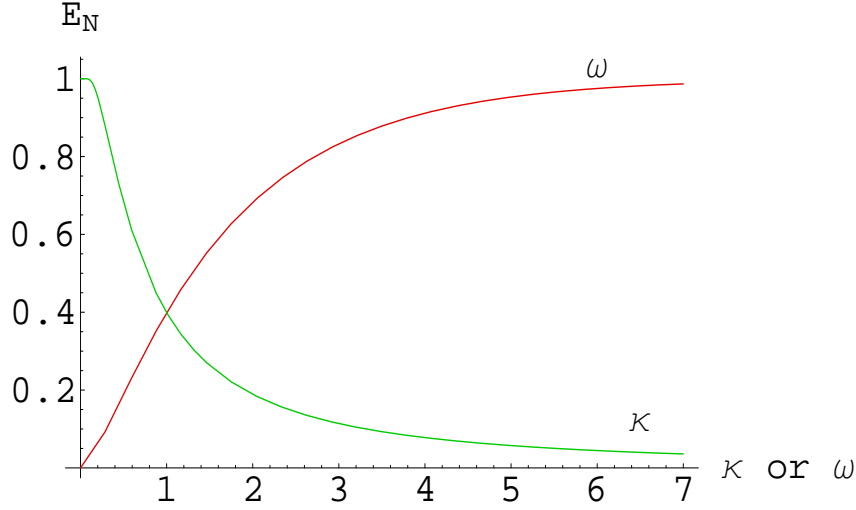 ,height=3in,width=4.5in }\caption{The
logarithmic negativity as a function of the surface gravity $\kappa$
and the mode frequency $\omega$.}
\end{figure}
\begin{figure}
\psfig{file=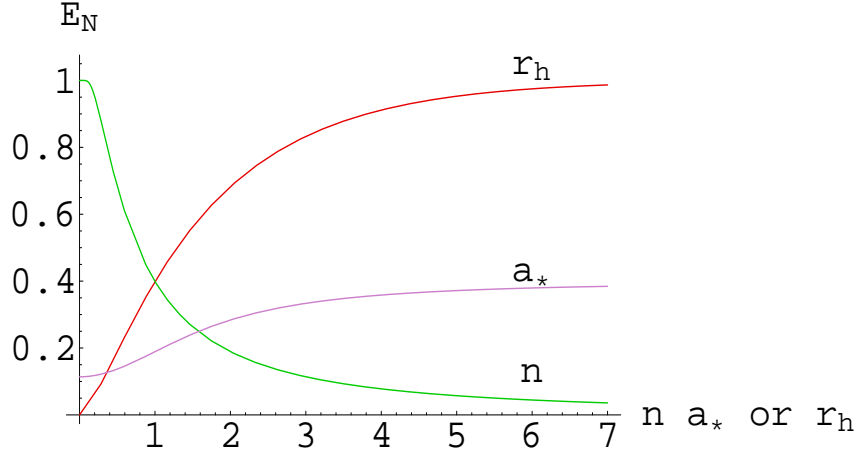 ,height=3in,width=4.5in }\caption{The
logarithmic negativity as a function of the horizon radius $r_{h}$,
the angular momentum $a_{*}$, and the extra dimensions $n$.}
\end{figure}
The analytic properties of this entanglement monotone are shown in
Fig. 5 and Fig. 6.  Figure 5 demonstrates that when the surface
gravity increases and the mode frequency $\omega$ is fixed, the
logarithmic negativity decreases. But if we fix the surface gravity
$\kappa$ and change the frequency $\omega$, the logarithmic
negativity increases with $\omega$.

Figure 6 shows that if we fix the angular momentum parameter $a$ and
vary the mass $\mu$ (or the horizon radius), the logarithmic
negativity increases as the mass (or the horizon radius) increases.
If we fix the black hole mass and vary the angular momentum
parameter $a$, the logarithmic negativity increases with $a_{*}$,
but the correction of entanglement monotone is small due to the
different value of $a_{*}$. This is different from a 4-dimensional
Kerr black hole \cite{ge2}. On the other hand, if we keep the mass
and the angular momentum parameter $a_{*}$ fixed, but vary the extra
dimensions of spacetime, the logarithmic negativity is reduced as
$n$ increases.

The quantum teleportation fidelity between Alice and Bob can also be
described by Eq. (\ref{fidelity}), that is to say,
\begin{equation}
\label{fid}
  F^{I}(\vert \varphi \rangle)=(1-e^{-\pi
(\omega-\omega_{0})/\kappa})^{3}.
\end{equation}
Figure 7 shows that the fidelity of teleportation is closely related
to the mode frequency and the surface gravity. Increasing the
surface gravity reduces the value of fidelity, while the mode
frequency does not.
\begin{figure}
\psfig{file=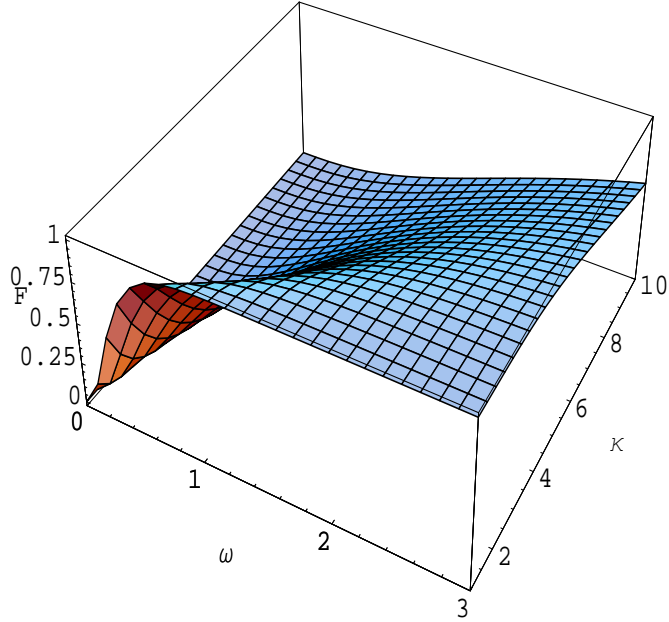 ,height=3.5in,width=3.5in }\caption{The
fidelity of teleportation with bosonic particles for a
$(4+n)$-dimensional Kerr black hole as a function of the mode
frequency $\omega$ and surface gravity $\kappa$.}
\end{figure}

\subsection{Quantum entanglement and teleportation with Dirac particles}

It is interesting to study quantum entanglement and teleportation
with fermions, since the Pauli exclusion principle becomes important
in this case. The 1-excitation Minkowski Fock state can be obtained
by acting the operator $a^{\dagger}_{\rm M}=\cos r
c^{\dagger}_{I}+\sin r c_{II}$ on the  Minkowski vacuum $|0>_{\rm
M}$, which reads as
\begin{equation}
\vert 1 \rangle_{\rm M}^{{\mathcal{~B}}}= \vert 1\rangle_{\rm I}
\otimes \vert 0\rangle_{\rm II}.
\end{equation}
Thus, we rewrite Eq. (\ref{bell}) in terms of Minkowski modes for
Alice and Kerr modes  for Bob. We trace over region $II$  and obtain
the resulting state
\begin{eqnarray}
&&\rho_{\rm AB}=\frac{1}{2}\left(\cos^2 r \vert 00 \rangle \langle
00 \vert +\sin^2 r \vert 01\rangle \langle 01 \vert  \right
 . \nonumber\\
 &&\left .
+ \cos r \vert 00 \rangle \langle 11 \vert + \cos r \vert 11 \rangle
\langle 00 \vert + \vert 11 \rangle \langle 11 \vert \right).
\end{eqnarray}
\begin{figure}
\psfig{file=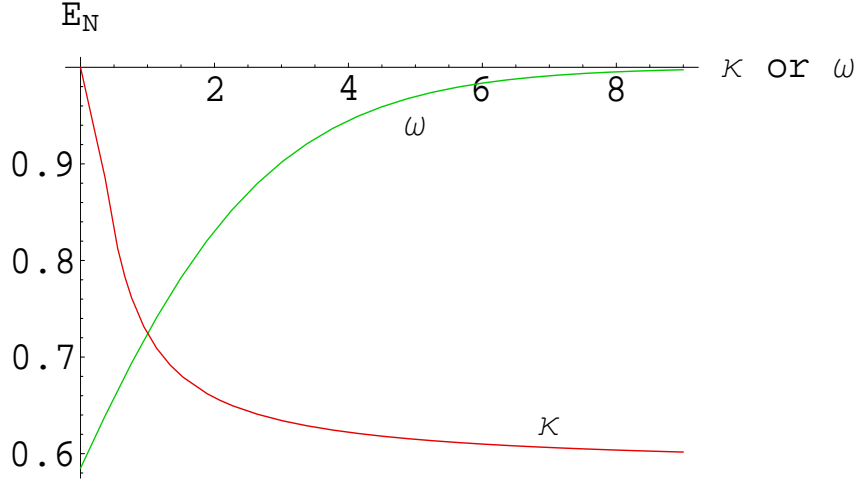 ,height=3in,width=4.5in }\caption{Logatihmic
negativity as a function of the surface gravity $\kappa$ and the
mode frequency $\omega$. Note that here $E_{\mathcal{N}\rm
min}=0.58$}.
\end{figure}
\begin{figure}
\psfig{file=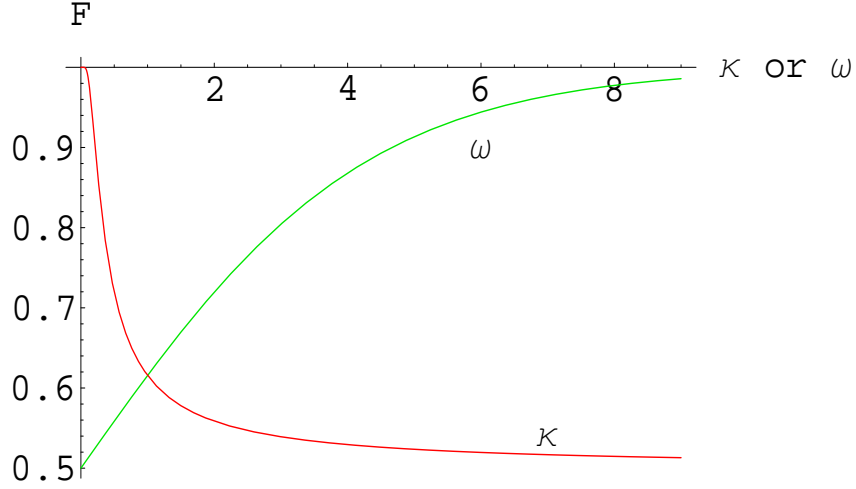 ,height=3in,width=4.5in }\caption{The
fidelity of teleportation is degraded with the surface gravity
$\kappa$ but is bounded to $1/2$ as $\kappa \rightarrow \infty$. }
\end{figure}
\begin{figure}
\psfig{file=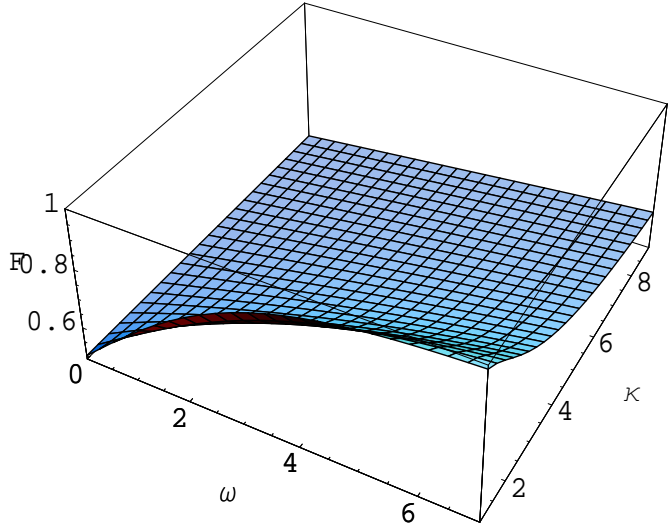 ,height=3in,width=3.5in }\caption{The
fidelity of teleportation with Dirac particles as a function of the
surface gravity and mode frequency. }
\end{figure}
The partial transpose for $\rho_{\rm AB}$ is obtained by
interchanging Alice's qubit as
\[
\rho_{AB}^{T}=\frac{1}{2}\left(
\begin{array}{cccc}
\cos^2 r & 0 & 0 & 0\\ 0 & \sin^2 r & \cos r & 0\\0 & \cos r & 0 &
0\\0 & 0& 0 & 1
\end{array}
\right),
\]
which yields one negative eigenvalue
\begin{equation}
\lambda_{1}=-\cos^2 r/2.
\end{equation}
The logarithmic negativity is found to be
\begin{equation}
\label{roe} E_{\mathcal{N}} = \log_{2}\left(1+\cos^2 r\right)
\end{equation}
>From Eq. (\ref{roe}), we can see that when the surface gravity
$\kappa$ approaches infinity, the logarithmic negativity saturates
to $\lim_{\kappa\rightarrow \infty}E_{\mathcal{N}}\simeq 0.58$ (see
Fig. 8).

We now turn to teleportation between Alice and Bob and utilize dual
rail basis states as an excitation of a spin-up state in one of two
possible spatial modes in Alice's cavity and similarly for Bob. Upon
receiving Alice measurement results, Bob trace over region $II$ of
the state he observes and obtains \cite{al2},
\begin{eqnarray}
\rho^{I}&=&\sum^{1}_{k=0}\sum^{1}_{l=0}{}_{I} \langle k,l\vert
\phi_{ij} \rangle \langle \phi_{ij}\vert
 k,l\rangle_{I}\nonumber\\
 &=& \cos^2 r \vert \phi_{ij}\rangle \langle \phi_{ij} \vert
 +\sin^2 r \vert 11 \rangle_{I}\langle
 11 \vert.
\end{eqnarray}
The fidelity of Bob's final state is then
\begin{equation}
F^{I}=Tr_{I}\left(\vert \psi \rangle \langle \psi \vert \rho^{I}
\right)
 = \cos^2 r.
\end{equation}
In the fermionic case, the number of allowed excitations is bounded
above by $n=2$. Thus when the acceleration approaches infinity, the
fidelity saturates to $\lim_{\kappa\rightarrow \infty}\cos^2 r=1/2$
(see Fig.9). Figure 10 demonstrates the fidelity of teleportation as
a function of both the mode frequency and black hole surface
gravity. For a mini-black hole that may be created at the LHC with
emitted Hawking radiation frequency range  $0.05\lesssim \omega
r_{h} \lesssim 0.5$, the logarithmic negativity for entanglement is
then bounded to $E_{\mathcal{N}\rm max}=0.73$ and the teleportation
saturates to $F_{\rm max}=0.65$.

\section{discussions and conclusions}

In summary, we have investigated quantum entanglement and
teleportation in the spacetime of higher dimensional Schwarzschild
and Kerr black holes. The teleportation protocols in the black hole
spacetime are discussed in detail. For massive black holes and
well-insulated cavities, which can reflect all non-gravitational
fields, it is possible for an observer with his cavity  to lower the
cavity slowly down to the black hole horizon, and perfect
entanglement and teleportation can be realized since the
gravitational coupling is negligible. However, for mini-black holes,
we prefer to let one observer (Bob) fall freely, then slow down and
stop to become stationary at a fixed radius. Actually, it is hard to
believe that we can keep a   thick wall cavity near a mini-black
hole horizon in thermal equilibrium with the Hawking radiation. If
we want the the cavity to be in thermal equilibrium, the wall of the
cavity should be very thin (compared to the size of the horizon
curvature) and a thin wall cavity could not prevent outside
non-gravitational fields from leaking into the cavity. We have
discussed a possible new physics that may arise from the presence of
extra dimensions and the unbounded angular momentum parameter
$a_{*}$. It is found that if the gravity is really at TeV scales,
the reduction of entanglement monotone and teleportation fidelity is
not so drastic as that in Planck scale gravity theory for mini-black
holes. We can use quantum entanglement and quantum teleportation
experiments to detect extra dimensions. For instance, for a
mini-black hole created at LHC, quantum entanglement experiments can
be conducted near this black hole. If this black hole is a higher
dimensional one, the reduction of entanglement monotone is not so
drastic as that of an ordinary 4-dimensional one. The reduction of
entanglement monotone can be described as a function of extra
dimensions. If we fix other parameters, then the degree of
entanglement decreases as the number of extra dimensions increases.

The degree of entanglement is found to be degraded with increasing
the extra dimensions. For a finite black hole surface gravity, the
observer may choose higher frequency mode to keep high level
entanglement. The fidelity of quantum teleportation is also reduced
because of the Hawking-Unruh effect. We have discussed the fidelity
as a function of the extra dimensions, mode frequency, mass and/or
angular momentum parameter of black hole for both bosonic and
fermionic resources. For quantum entanglement and teleportation with
bosonic particles, all the correlations might be lost for the
observers in gravitational fields and the fidelity could be degraded
to zero, but for fermionic cases, the entanglement monotone and
teleportation fidelity are bounded to $0.58$ and $0.5$,
respectively, because of the Pauli exclusion principle.\\

\acknowledgements The authors would like to thank V.~Frolov,
R.~Mann, W.~T.~Ni and D.~N.~Page for their helpful comments. The
work of S.~P.~K. was supported by the Korea Science and Engineering
Foundation (KOSEF) grant funded by the Korea government (MOST) (No.
F01-2007-000-10188-0).

\end{document}